\pdfoutput=1

\documentclass[3p,times,twocolumn]{elsarticle}

\usepackage{amssymb}
\usepackage{amsmath}
\usepackage[figuresright]{rotating}

\makeatletter
\def\ps@pprintTitle{%
  \let\@oddhead\@empty
  \let\@evenhead\@empty
  \def\@oddfoot{\reset@font\hfil\thepage\hfil}
  \let\@evenfoot\@oddfoot
}
\makeatother

\newcommand{\antinue}{\overline{\nu}_{e}}
\newcommand{\tot}{\theta_{13}}
\newcommand{\sstot}{\sin^{2}2\theta_{13}}

\newcommand{\californium}{\rm{{}^{252}Cf}}
\newcommand{\lithium}{\rm{{}^{9}Li}}
\newcommand{\helium}{\rm{{}^{8}He}}
\newcommand{\chisq}{\chi^{2}}
\newcommand{\MeV}{\text{\,MeV}}

\begin{document}

\begin{frontmatter}

\title{Double Chooz: Latest results}

\author{J.I. Crespo-Anad\'on}

\address{Centro de Investigaciones Energ\'eticas, Medioambientales y Tecnol\'ogicas, CIEMAT, 28040, Madrid,
Spain}

\begin{abstract}
The latest results from the Double Chooz experiment on the neutrino mixing angle $\theta_{13}$ are presented.
A detector located at an average distance of 1050 m from the two reactor cores of the Chooz nuclear power plant has accumulated a live time of 467.90 days, corresponding to an exposure of 66.5 GW-ton-year (reactor power $\times$ detector mass $\times$ live time).
A revised analysis has boosted the signal efficiency and reduced the backgrounds and systematic uncertainties compared to previous publications, paving the way for the two detector phase.
The measured $\sin^2 2\theta_{13} = 0.090^{+0.032}_{-0.029}$ is extracted from a fit to the energy spectrum. A deviation from the prediction above a visible energy of 4 MeV is found, being consistent with an unaccounted reactor flux effect, which does not affect the $\theta_{13}$ result.
A consistent value of $\tot$ is measured in a rate-only fit to the number of observed candidates as a function of the reactor power, confirming the robustness of the result.
\end{abstract}


\end{frontmatter}

\section{Introduction}
\label{sec:Introduction}
Neutrino oscillations in the standard three-flavor framework are described by three mixing angles, three mass-squared differences (two of which are independent) and one CP-violating phase. Excepting the phase which still remains unknown, all the other parameters have been measured \cite{PDG2014}. $\tot$ was the last to be measured by short-baseline reactor and long-baseline accelerator experiments \cite{DC2ndPub, DCHPub, DCRRMElsevier, DC3rdPub, DayaBayShape, RENO, MINOSTheta13, T2KTheta13}.   

For the energies and distances relevant to Double Chooz, the oscillation probability is well approximated by the two-flavor case. Thus, the survival probability reads:
\begin{equation}
 P_{\antinue \to \antinue} = 1 - \sin^2 2\tot \sin^2 \left( 1.27 \frac{\Delta m^2_{31} [\text{eV}^{2}] L [\text{m}]}{E_{\nu}[\text{MeV]}} \right)
\end{equation} 
So $\tot$ can be measured from the deficit in the electron antineutrino flux emitted by the reactors. In this analysis, $\Delta m^2_{31} = 2.44^{+0.09}_{-0.10} \times 10^{-3} \text{\,eV}^2$, taken from \cite{MINOSDM231}, assuming normal hierarchy . 

Antineutrinos are detected through the inverse beta-decay (IBD) process on protons, $\antinue + p \to e^+ + n$, which provides two signals: a prompt signal in the range of 1 - 10 MeV is given by the positron kinetic energy and the resulting $\gamma$s from its annihilation. This visible energy is related to the $\antinue$ energy by $E_{vis} \approx E_{\nu} - 0.8 \text{\,MeV}$. A delayed signal is given by the $\gamma$s released in the radiative capture of the neutron by a Gd or H nucleus. The results presented here correspond only to captures in Gd, which occur after a mean time of 31.1 $\mu$s and release a total energy of 8 MeV, which is far above the natural radioactivity energies. The coincidence of these two signals grants the experiment a powerful background suppression. 

\section{The Double Chooz experiment}
\label{sec:DoubleChooz}
Double Chooz (DC) is a 2-detector experiment located in the surroundings of the Chooz nuclear power plant (France), which has two pressurized water reactor cores, producing 4.25 $\rm{GW_{th}}$ each. The Near Detector (ND), placed at $\sim 400$ m from the cores, has a 120 m.w.e. overburden and it is currently being commissioned. The Far Detector (FD), placed at $\sim 1050$ m from the cores, has a 300 m.w.e. overburden and its data are used here. The 2-detector concept allows to extract $\tot$ with high precision from the relative comparison of the $\antinue$ flux at the two detectors. Because the detectors are built identical, all the correlated uncertainties between them are cancelled.

Since the ND was not operative for this analysis yet, an accurate reactor flux simulation was needed to obtain the $\antinue$ prediction. \'Electricit\'e de France provides the instantaneous thermal power of each reactor, and the location and initial composition of the reactor fuel. The simulation of the evolution of the fission rates and the associated uncertainties is done with \texttt{MURE} \cite{MURE1, MURE2}, which has been benchmarked with another code \cite{DRAGON}. The reference $\antinue $ spectra for $\rm{{}^{235}U}$, $\rm{{}^{239}Pu}$ and $\rm{{}^{241}Pu}$ are computed from their $\beta$ spectrum \cite{ILL1, ILL2, ILL3}, while \cite{Haag238U} is used for $\rm{{}^{238}U}$ for the first time. The short-baseline Bugey4 $\antinue$ rate measurement \cite{Bugey4} is used to suppress the normalization uncertainty on the $\antinue$ prediction, correcting for the different fuel composition in the two experiments. The systematic uncertainty on the $\antinue$ rate amounts to 1.7\%, dominated by the 1.4\% of the Bugey4 measurement. Had the Bugey4 measurement not been included, the uncertainty would have been 2.8\%.

\begin{figure}[ht]
\begin{center}
	\includegraphics[trim=15cm 30cm 0cm 20cm,clip=true, width=0.5\textwidth]{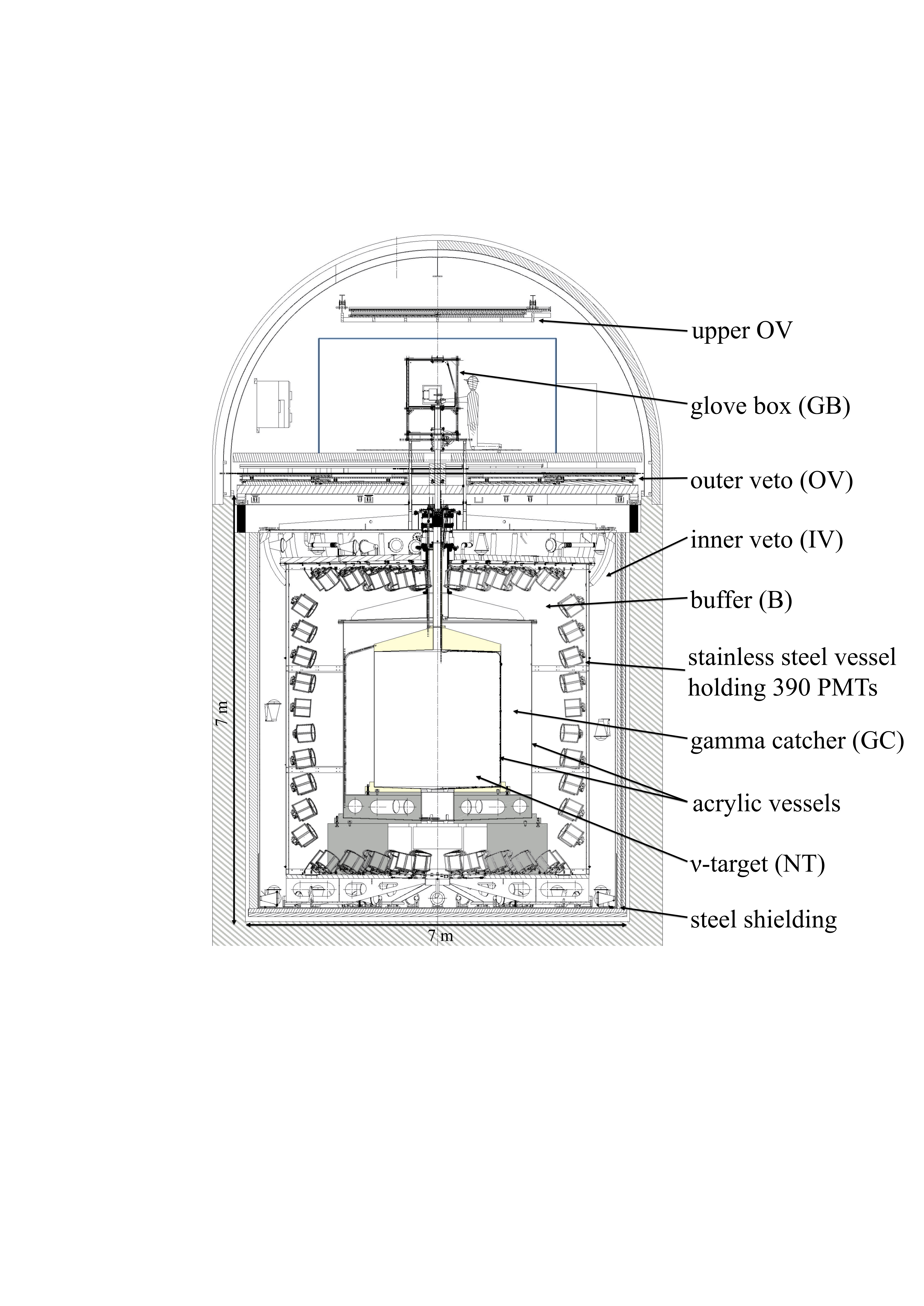}
	\caption{Double Chooz far detector design.}
	\label{fig:Detector}
\end{center}
\end{figure}

The DC detector is composed of four concentrical cylindrical vessels (see figure \ref{fig:Detector}). The innermost volume, the $\nu$-target (NT), is an 8 mm thick acrylic vessel (UV to visible transparent) filled with 10.3 $\rm{m^3}$ of liquid scintillator loaded with Gd (1g/l) to enhance the neutron captures.
The $\gamma$-catcher (GC), a 55 cm thick layer of liquid scintillator (Gd-free) enclosed in a 12 mm thick acrylic vessel surrounds the NT to maximize the energy containment. Surrounding the GC is the buffer, a 105 cm thick layer of mineral oil (non-scintillating) contained in a stainless steel tank where 390 low background 10-inch photomultiplier tubes (PMT) are installed, and which shields from the radioactivity of the PMTs and the surrounding rock. The elements described so far constitute the inner detector (ID). Enclosing the ID and optically separated from it, the inner veto (IV), a 50 cm thick layer of liquid scintillator, serves as a cosmic muon veto and as an active shield to incoming fast neutrons observed by 78 8-inch PMTs positioned on its walls. A 15 cm thick demagnetized steel shield protects the whole detector from external $\gamma$-rays. The outer veto (OV), two orthogonally aligned layers of plastic scintillator strips placed on top of the detector, allows a 2D reconstruction of impinging muons. An upper OV covers the chimney, which is used for filling the volumes and for the insertion of calibration sources (encapsulated radioactive sources of $^{137}$Cs, $^{68}$Ge, $^{60}$Co and $\californium$ and a laser). Attached to the ID and IV PMTs, a multi-wavelength LED-fiber light injection system is used to periodically calibrate the readout electronics.

Waveforms from all ID and IV PMTs are digitized and recorded by dead-time free flash-ADC electronics.

DC has pioneered the measurement of $\tot$ using the $\antinue $ spectral information because of its exhaustive treatment of the energy scale, which is applied in parallel to the recorded data and the Monte Carlo (MC) simulation. 
A linearized photoelectron (PE) calibration produces a PE number in each PMT which has been corrected from dependencies on the gain non-linearity and time.
A uniformity calibration corrects for the spatial dependence of the PE, equalizing the response within the detector.
The conversion from PE to energy units is obtained from the analysis of neutron captures in H from a $\californium$ calibration source deployed at the center of the detector.
A stability calibration is applied to the data to remove the remaining time variation by analyzing the evolution of the H capture peak from spallation neutrons, which is also crosschecked at different energies using the Gd capture peak and the  $\alpha$ decays of $\rm{{}^{212}Po}$.
Two further calibrations are applied to the MC to correct for the energy non-linearity relative to the data: the first is applied to every event and it arises from the modeling of the readout systems and the charge integration algorithm; the second, which is only applied to positrons, is associated to the scintillator modeling.  
The total systematic uncertainty in the energy scale amounts to 0.74\%, improving the previous one \cite{DC2ndPub} by a 1.5 factor.

\section{Neutrino selection}
\label{sec:Selection}
The minimum energy for a selected event is $E_{vis} > 0.4 \text{\,MeV}$, where the trigger is already 100\% efficient.
Events with $E_{vis} > 20 \text{\,MeV}$ or $E_{IV} > 16 \text{\,MeV}$ are rejected and tagged as muons, imposing a 1 ms veto after them to reject also muon-induced events.
\textit{Light noise} is a background caused by spontaneous light emission from some PMT bases, and it is avoided by requiring the selected events to satisfy all the following cuts: i) $q_{max}$, the maximum charge recorded by a PMT, must be less or equal to 12\% of the total charge of the event; ii)  $1/N \times \sum\nolimits_{i = 0}^{N} (q_{max} - q_i)^2/q_i < 3 \times 10^4 \text{\,charge units}$, where $N$ is the number of PMTs located at less than 1 m from the PMT with the maximum charge; iii) $\sigma_t < 36 \text{\,ns}$ or $\sigma_q > (464 - 8\sigma_t)\text{\, charge units}$, where $\sigma_t$ and $\sigma_q$ are the standard deviations of PMTs hit times and integrated charge distributions, respectively.
Events passing the previous cuts are used to search for coincidences, which must satisfy the conditions: the prompt $E_{vis}$ must be in $(0.5, 20)\MeV$, the delayed $E_{vis}$ in $(4, 10) \MeV$, the correlation time between the signals must be in $(0.5, 150) \,\mu$s, and the distance between reconstructed vertex positions must be less than 1 m. In addition, only the delayed signal can be in a time window spanning 200 $\mu$s before and 600 $\mu$s after the prompt signal.

\section{Background measurement and vetoes}
\label{sec:Background}
The backgrounds are non-neutrino processes which mimic the characteristic coincidence of the IBD.

\paragraph{Cosmogenic isotopes}
Unstable isotopes are produced by spallation of nuclei inside the detector by cosmic muons. Products as $\lithium$ and $\helium$ have a decay mode in which a neutron is emitted along with an electron, indistinguishable from an IBD interaction. Moreover, lifetimes of $\lithium$ and $\helium$ are 257 ms and 172 ms, respectively, so the 1 ms after-muon veto is not effective.
A cut in a likelihood based on the event distance to the muon track and the number of neutron candidates following the muon in 1 ms allows to reject 55\% of $\lithium$ and $\helium$.
The $\lithium/\helium$ contamination is determined from fits to the time correlation between the IBD candidates and the previous muon. The estimation of the remaining $\lithium/\helium$ background in the IBD candidates sample is $0.97^{+0.41}_{-0.16}$ events/day. The events vetoed by the likelihood cut are used to build the prompt energy spectrum (see figure \ref{fig:BGRSFit}), which includes also captures on H to enhance the statistics.

\paragraph{Fast neutrons and stopping muons}
Fast neutrons originated from spallation by muons in the surrounding rock can enter the detector and reproduce the IBD signature by producing recoil protons (prompt) and be captured later (delayed). Stopping muons are muons which stop inside the detector, giving the prompt signal, and then decay producing the Michel electron that fakes the delayed signal.
In order to reject this background, events fulfilling at least one of the following conditions are discarded:
(i) Events with an OV trigger coincident with the prompt signal.
(ii) Events whose delayed signal is not consistent with a point-like vertex inside the detector.
(iii) Events in which the IV shows correlated activity to the prompt signal.
The three vetoes together reject 90\% of the events with a prompt $E_{vis} > 12 \MeV$, where this background is dominant.
The veto (iii) is used to extract the fast neutron/stopping muon prompt energy spectrum, which is found to be flat. This shape is further confirmed by using the other vetoes. The rate of this background in the candidate sample is estimated from a IBD-like coincidence search in which the prompt signal has an energy in the $(20, 30)\MeV$ region, and it amounts to $0.604\pm 0.051$ events/day.

\paragraph{Accidental background}
Those are random coincidences of two triggers satisfying the selection criteria. Because of its random nature, their rate and spectrum (see figure \ref{fig:BGRSFit}) can be studied with great precision from the data by an off-time coincidence search, the same as the IBD selection except for the correlated time window, which is opened more than 1 s after the prompt signal. The use of multiple windows allows to collect high statistics. The background rate is measured to be $0.0701 \pm 0.0003 \text{\,(stat)} \pm 0.026 \text{\,(syst)}$.

Other backgrounds, such as the $^{13}$C($\alpha$, n)$^{16}$O reaction or the $^{12}$B decay, were considered but they were found to have negligible occurrence. Table \ref{table:BGSummary} summarizes the estimated background rates and the reduction with respect to the previous publication \cite{DC2ndPub}.

\begin{table}[h]
\begin{center}
	\begin{tabular}{ccc}
	Background & Rate (d$^{-1}$)& \cite{DC3rdPub}/\cite{DC2ndPub}\\
	\hline
	$^{9}$Li/$^{8}$He & $0.97^{+0.41}_{-0.16}$ & 0.78\\
	Fast-n/stop-$\mu$ & $0.604 \pm 0.051$ & 0.52 \\
	Accidental & $0.070 \pm 0.003$ & 0.27 \\
         $^{13}$C($\alpha$, n)$^{16}$O & $< 0.1$ & N/A in \cite{DC2ndPub}\\
         $^{12}$B & $< 0.03$ & N/A in \cite{DC2ndPub} \\
	\end{tabular}
	\caption{Summary of background rate estimations. \cite{DC3rdPub}/\cite{DC2ndPub} shows the reduction of the background rate in \cite{DC3rdPub} with respect to the previous publication \cite{DC2ndPub}, after correcting for the different prompt energy range.}
	\label{table:BGSummary}
\end{center}
\end{table}

\section{IBD detection efficiency}
\label{sec:Efficiency}
A dedicated effort was carried out to decrease the detection efficiency uncertainty. This signal normalization uncertainty is dominated by the neutron detection uncertainty, which has been reduced from 0.96\% in \cite{DC2ndPub} to the current 0.54\% in \cite{DC3rdPub}. This was achieved thanks to the reduction of the volume-wise selection systematic uncertainty by using two new methods to estimate the neutron detection efficiency in the full Target. The first one uses the neutrons produced by the IBD interactions, which are homogeneously distributed in the detector, to produce a direct measurement of the volume-wide efficiency. The second method exploits the symmetry shown by the neutron detection efficiency, in which the data from the $\californium$ source deployed along the vertical coordinate can be extrapolated to the radial coordinate. Another reduction was obtained on the uncertainty arising from the spill-in/spill-out currents (neutron migration into and out of the NT, respectively), which are sensitive to the low energy neutron physics. It was decreased by comparing the custom DC \texttt{Geant4} simulation, which includes an analytical modeling of the impact of the molecular bonds on low energy neutrons, to \texttt{Tripoli4}, a MC code with a specially accurate model of low energy neutron physics.

After accounting for the uncertainties introduced by the background vetoes and the scintillator proton number, the detection-related normalization uncertainty totals 0.6\%.

\section{Oscillation analyses}
\label{sec:Fit}
In a live-time of 460.67 days with at least one reactor running, 17351 IBD candidates were observed. The prediction, including backgrounds, in case of no oscillation was $18290^{+370}_{-330}$. The deficit is understood as a consequence of neutrino oscillation. In addition, a live-time of 7.24 days with the two reactors off was collected \cite{DCOffOff}, in which 7 IBD candidates were observed, whereas the prediction including the residual $\antinue$ was $12.9^{+3.1}_{-1.4}$. The reactor-off measurement allows to test the background model and constrain the total background rate in the oscillation analysis. It is a unique advantage of DC, which has only two reactors.

The normalization uncertainties of the signal and the background are summarized in table \ref{table:RateError}, showing also the improvement with respect to the previous analysis \cite{DC2ndPub}.

\begin{table}[h]
\begin{center}
	\begin{tabular}{ccc}
	Source & Uncertainty (\%) & \cite{DC3rdPub}/\cite{DC2ndPub}\\
	\hline
	Reactor flux & 1.7 & 1.0\\
	Detection efficiency & 0.6 & 0.6 \\
	$^{9}$Li/$^{8}$He & $+1.1$ / $-0.4$ & 0.5\\
	Fast-n/stop-$\mu$ & 0.1 & 0.2\\
	Statistics & 0.8 & 0.7\\
	\hline
	Total & $+ 2.3$ / $-2.0$ & 0.8 \\
	\end{tabular}
	\caption{Signal and background normalization uncertainties relative to the signal prediction. \cite{DC3rdPub}/\cite{DC2ndPub} shows the reduction of the uncertainty with respect to the previous publication \cite{DC2ndPub}.}
	\label{table:RateError}
\end{center}
\end{table}

\subsection{Reactor rate modulation analysis}
\label{subsec:RRMFit}

\begin{figure}[h]
\begin{center}
	\includegraphics[width=0.45\textwidth]{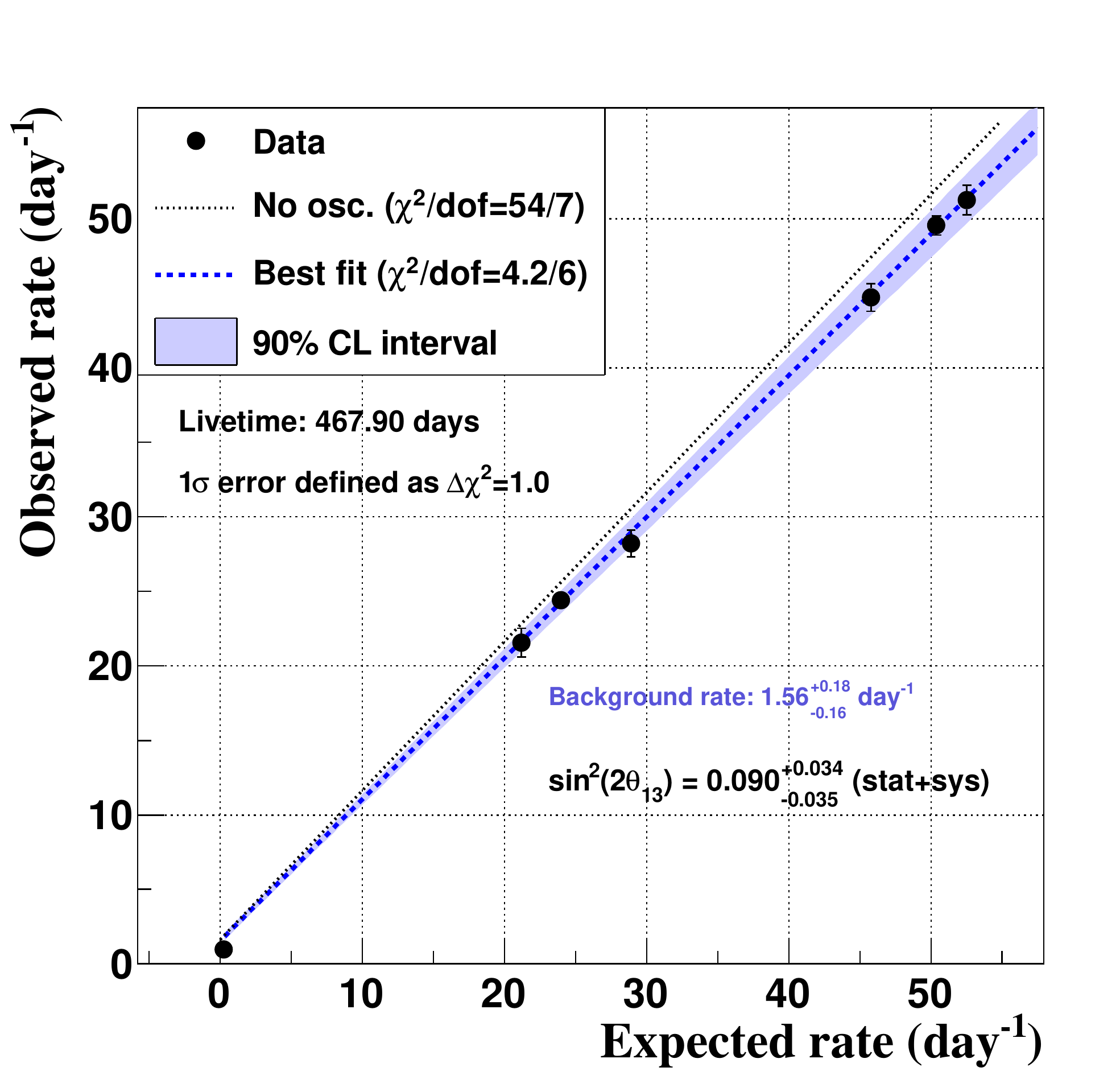}
	\caption{Observed versus expected candidate daily rates for different reactor powers.The prediction under the null oscillation hypothesis (dotted line) and the best-fit with the background rate constrained by its uncertainty (blue dashed line) are shown. The first point corresponds to the reactor-off data.}
	\label{fig:RRMFit}
\end{center}
\end{figure}

From the linear correlation existing between the observed and the expected candidate rates at different reactor conditions, a fit to a straight line determines simultaneously $\sstot$ (proportional to the slope) and the total background rate $B$ (intercept) \cite{DCRRMElsevier}. Including the prediction of the total background $B = 1.64^{+0.41}_{-0.17} \text{\,events/day}$, the best fit is found at $\sstot = 0.090^{+0.034}_{-0.035}$ and $B = 1.56^{+0.18}_{-0.16} \text{\,events/day}$ (see figure \ref{fig:RRMFit}).

A background model independent measurement of $\tot$ is possible when the background constraint is removed and $B$ is treated as a free parameter. The best fit ($\chi^2_{min}/d.o.f. = 1.9/5$) corresponds to $\sstot = 0.060 \pm 0.039$ and $B = 0.93^{+0.43}_{-0.36} \text{\,events/day}$, consistent with the background-constrained fit.

The impact of the reactor-off data is tested by removing the reactor-off point (with the background rate still unconstrained). In this case, the best fit ($\chi^2_{min}/d.o.f. = 1.3/4$) gives $\sstot = 0.089 \pm 0.052$ and $B = 1.56 \pm 0.86 \text{\,events/day}$, which confirms the improvement granted by the reactor-off measurement. 

\subsection{Rate + shape analysis}
\label{subsec:RSFit}

\begin{figure}[hb]
\begin{center}
	\includegraphics[width=0.5\textwidth]{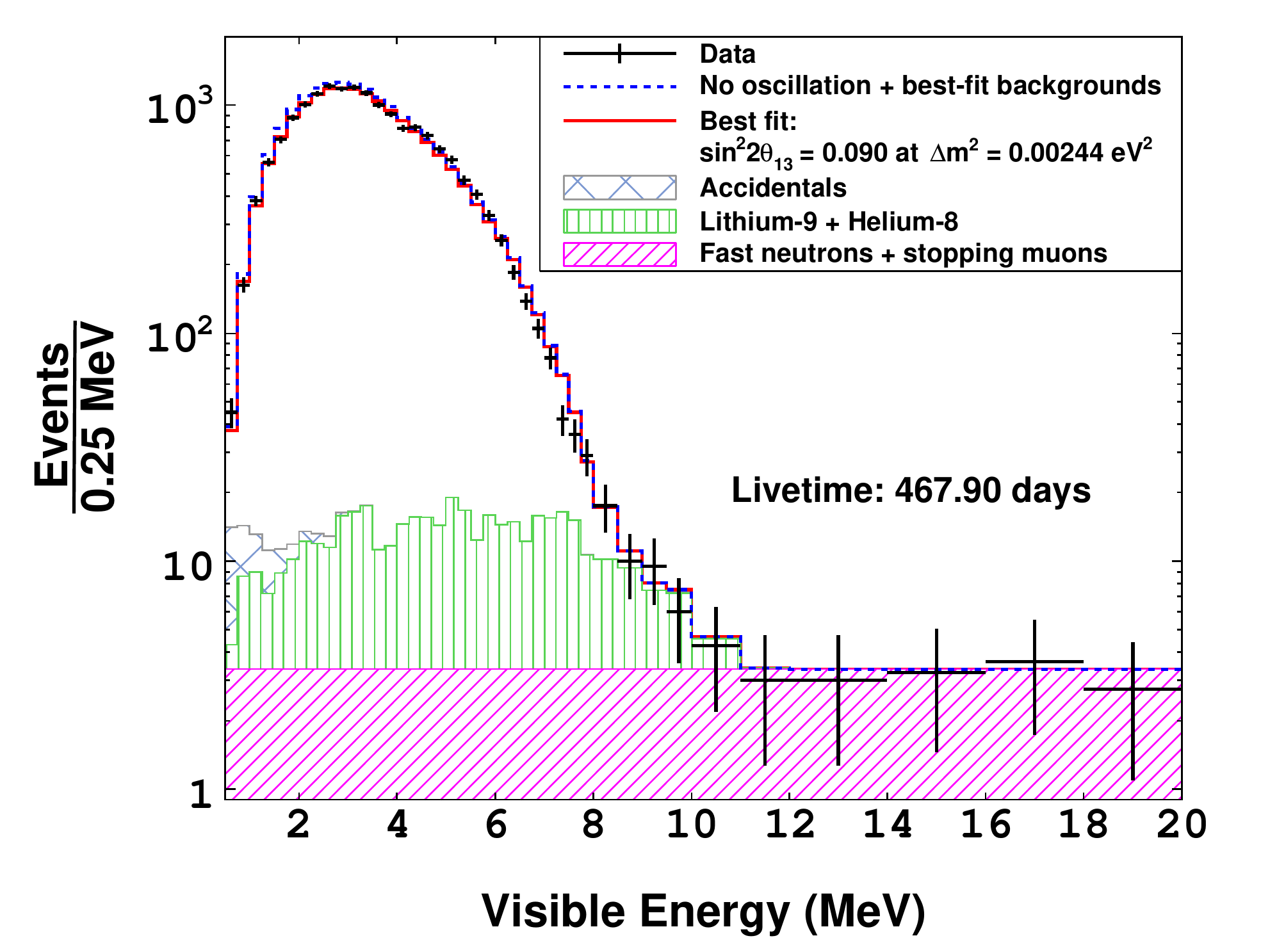}
	\caption{Measured prompt energy spectrum (black points with statistical error bars), superimposed on the no-oscillation prediction (blue dashed line) and on the best fit (red solid line), with the stacked best-fit backgrounds added.}
	\label{fig:BGRSFit}
\end{center}
\end{figure}

\begin{figure}[]
\begin{center}
	\includegraphics[width=0.5\textwidth]{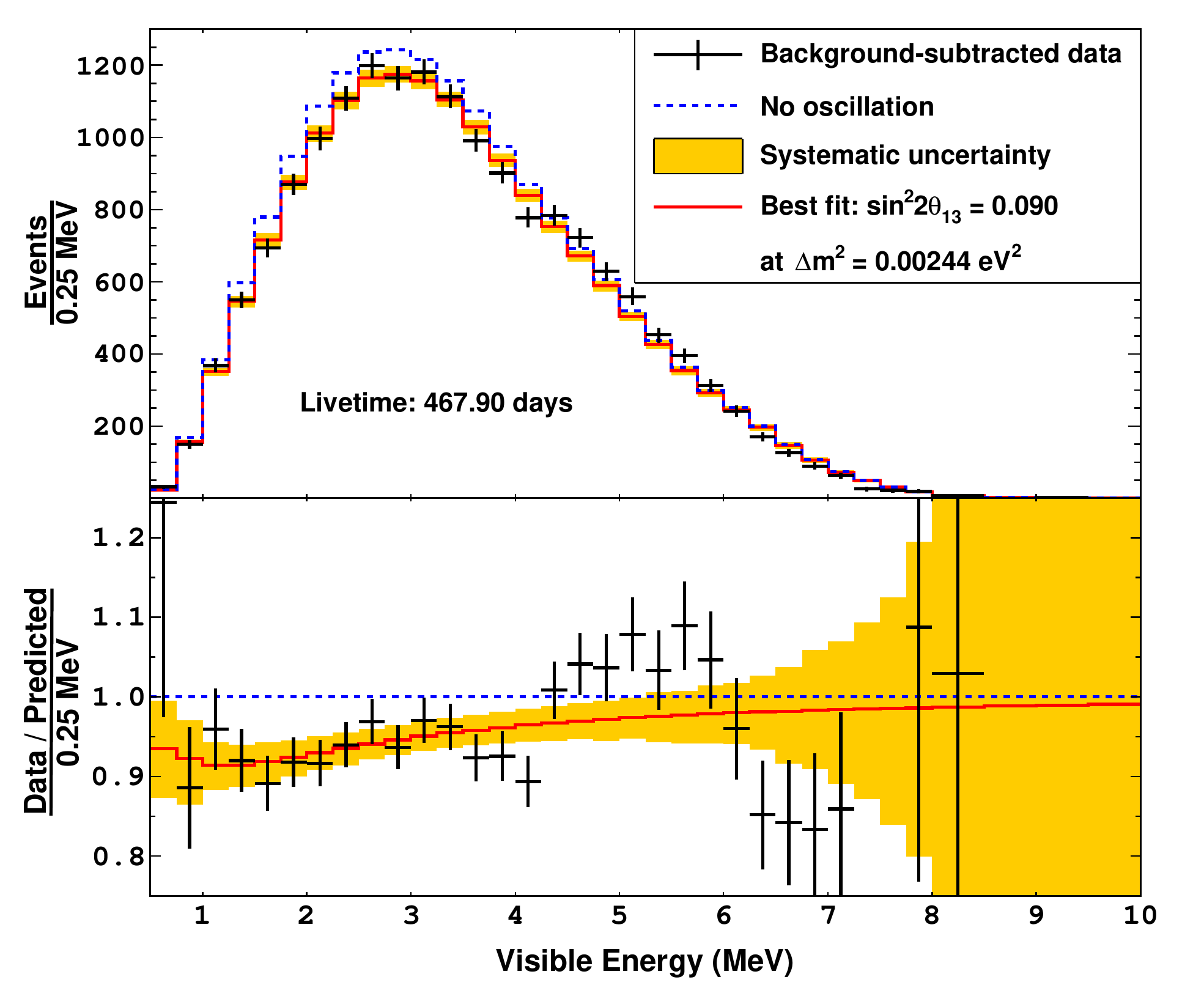}
	\caption{Top: Measured prompt energy spectrum with best-fit backgrounds subtracted (black points with statistical error bars) superimposed on the no-oscillation prediction (blue dashed line) and on the best fit (red solid line). Bottom: Ratio of data to the no-oscillation prediction (black points with statistical error bars) superimposed on the best fit ratio (red solid line). The gold band represents the systematic uncertainty on the best-fit prediction.}
	\label{fig:RSFit}
\end{center}
\end{figure}

This analysis measures $\sstot$ by minimizing a $\chisq$ in which the prompt energy spectrum of the observed IBD candidates and the prediction are compared. A covariance matrix accounts for the statistical and systematic (reactor flux, MC normalization, $\lithium/\rm{{}^8He}$ spectrum shape, accidental statistical) uncertainties in each bin and the bin-to-bin correlations. A set of nuisance parameters accounts for the other uncertainty sources: $\Delta m^2_{31}$, the number of residual $\antinue$ when reactors are off ($1.57 \pm 0.47$ events), the $\lithium/\rm{{}^8He}$ and fast neutron/stopping muon rates, the systematic component of the uncertainty on the accidental background rate, and the energy scale. The best fit ($\chi^2_{min}/d.o.f. = 52.2/40$) is found at $\sstot = 0.090^{+0.032}_{-0.029}$ (see figures \ref{fig:BGRSFit},\ref{fig:RSFit}).

In addition to the oscillation-induced deficit on the bottom panel of figure \ref{fig:RSFit}, a spectrum distortion is observed above 4 MeV. The excess has been found to be proportional to the reactor power, disfavoring a background origin. Considering only the IBD interaction, the structure is consistent with an unaccounted reactor $\antinue$ flux effect, which does not affect significantly the $\tot$. The good agreement with the shape-independent reactor rate modulation result demonstrates it. The existence of this distortion has been later confirmed by the Daya Bay and RENO reactor experiments.

\begin{figure}[]
\begin{center}
	\includegraphics[width=0.5\textwidth]{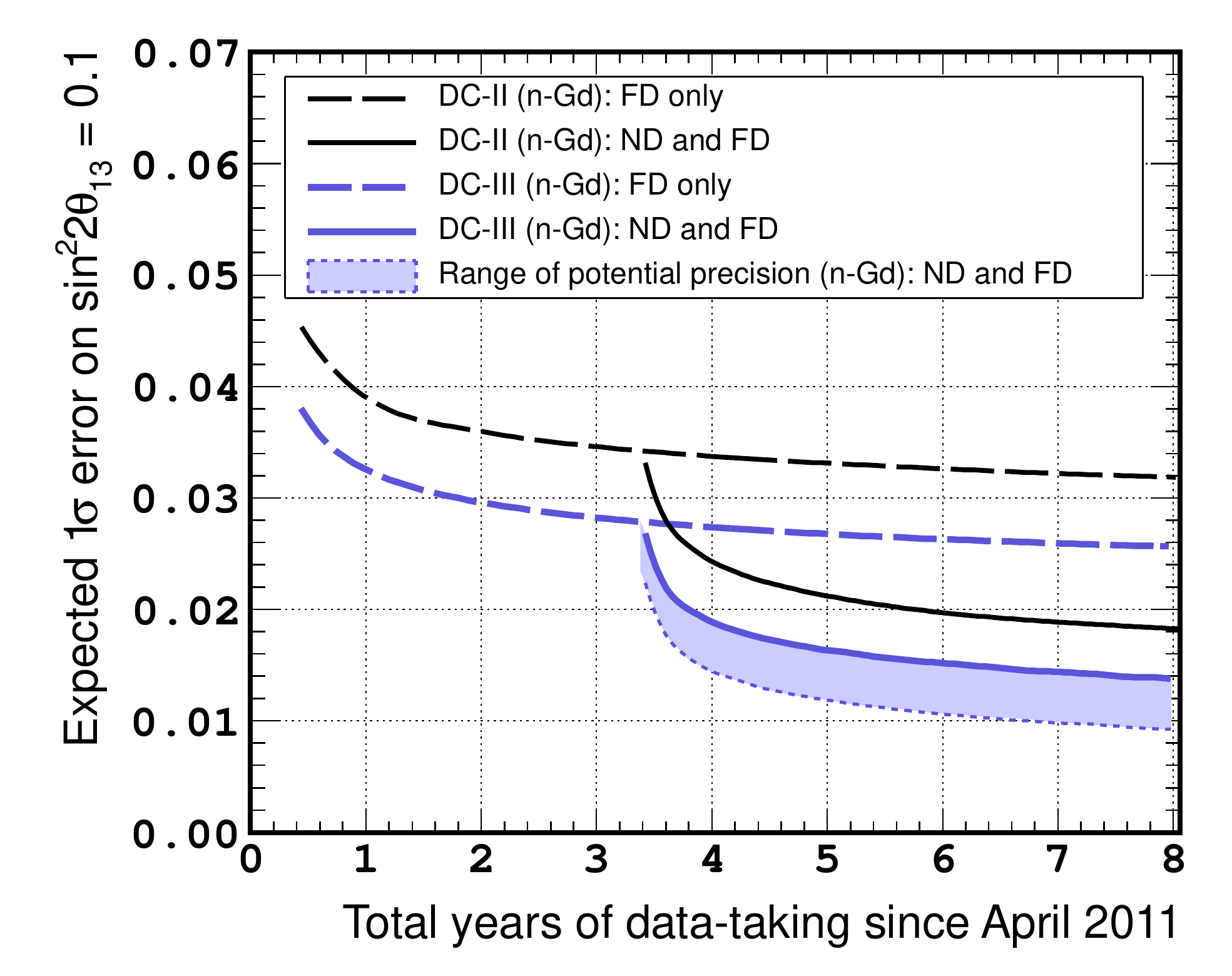}
	\caption{Double Chooz projected sensitivity using the IBD neutrons captured in Gd. The previous analysis, \cite{DC2ndPub} with only the FD (black dashed line) and adding the ND (black solid line), and the current analysis, with only the FD (blue dashed line) and adding the ND (blue solid line), are shown. The shaded region represents the range of improvement expected by reducing the systematic uncertainty, bounded from below by considering only the reactor systematic uncertainty.}
	\label{fig:sensitivity}
\end{center}
\end{figure}

Figure \ref{fig:sensitivity} shows the projected sensitivity of the Rate + Shape analysis using the IBD neutrons captured in Gd. A $0.2\%$ relative detection efficiency uncertainty is assumed, the expected remnant from the cancellation of the correlated detection uncertainties due to the use of identical detectors. The portion of reactor flux uncorrelated between detectors is $0.1\%$ (thanks to the simple experimental setup with two reactors). Backgrounds in the Near Detector are scaled from the Far Detector accounting for the different muon flux. Comparing the curves from the previous \cite{DC2ndPub} and the current analysis \cite{DC3rdPub}, the improvement gained with the new techniques is clear, and it is expected to improve further (e.g. systematic uncertainty on the background rate is limited by statistics).

\section{Conclusion}
\label{sec:Conclusion}

Double Chooz has presented improved measurements of $\tot$ corresponding to 467.90 days of live-time of a single detector using the neutrons captured in Gd. The most precise value is extracted from a fit to the observed positron energy spectrum: $\sstot = 0.090^{+0.032}_{-0.029}$. A consistent result is found by a fit to the observed candidate rates at different reactor powers: $\sstot = 0.090^{+0.034}_{-0.035}$. A distortion in the spectrum is observed above 4 MeV, with an excess correlated to the reactor power. It has no significant impact on the $\tot$ result.

As a result of the improved analysis techniques, Double Chooz will reach a 15\% precision on $\sstot$ in 3 years of data taking with two detectors, with the potential to improve to 10\%. 

\bibliographystyle{elsarticle-num}
\bibliography{DC}

\end{document}